\documentstyle[A4,12pt]{article}
\begin{document}
\newcommand{\be}{\begin{equation}}
\newcommand{\ee}{\end{equation}}
\newcommand{\bea}{\begin{eqnarray}}
\newcommand{\eea}{\end{eqnarray}}
\begin{titlepage}
\begin{flushright}
NBI-HE 96-61\\
hep-th/9610176\\
October 1996\\
\end{flushright}
\vspace*{0.5cm}
\begin{center}
{\bf
\begin{Large}
STRING SOLUTIONS TO SUPERGRAVITY\\
\end{Large}
}
\vspace*{3cm}
         {\large A. K. Tollst\'{e}n\footnote{email tollsten@nbi.dk}}
         \\[.5cm]
         The Niels Bohr Institute, Blegdamsvej 17,\\
         DK-2100 Copenhagen \O, Denmark
\end{center}
\vspace*{2cm}
\begin{abstract}
We find the comlete solution to ten-dimensional supergravity coupled 
to a three-form field strength, given the ``standard ansatz" for the fields,
and show that in addition to the well-known elementary and solitonic 
(heterotic) string solutions, one of the possibilities is an (unstable)
elementary type~I string solution.
\end{abstract}
\end{titlepage}

\section{Introduction}

A necessary condition for strong-weak coupling duality between two
theories is that the elementary states of one theory turn up as soliton-like
states in the other. This condition is satisfied for most of the 
conjectured dualities in 10 and 11-dimensional theories. However, an important 
exception is the heterotic-type~I duality in d=10,9,8 \cite{W95,Ts95,PW95}.
Here, the heterotic string does indeed turn up as a stable solution to
the low energy effective action of type~I theory \cite{Dab95,H195}, 
while there is no such
type~I solution to the low energy effective action of the heterotic
string. This is perhaps not very surprising. All other stable $p-1$-brane
solutions turning up in various dualities are {\em closed}, and, as we shall
see for the string case,  the solutions can be interpreted as a 
compactification (possibly with infinite compactification radii) of $p-1$
coordinates, around which the $p-1$-brane winds. Of course, no such topological
argument will hold in the open string case. On the contrary, a solution
consisting of an infinitely long open string with constant energy/length unit
is obviously unstable.

In this talk we review the string solutions to ten-dimensional supergravity.
We first find the general solution to the equations of motion given the 
``standard ansatz". These solutions make the equations of motion singular at
what can be interpreted as the location of a string source.
Requiring also this source to be dynamical takes us back to the well-known
supersymmetric fundamental string solution \cite{DGHR90}. 
In addition, there are other 
solutions, representing the field configuration around a fixed source. One
of these shows the correct behaviour under Weyl rescaling to be a type~I
solution. It has no conserved charge, nor does it preserve supersymmetry,
so there is no reason to believe that it is stable.

Further details of the calculation, notation and conventions
are published elsewhere \cite{T96}.

\section{The string solution of ten-dimensional supergravity}
We would like to study the bosonic part of the combined string supergravity
action in ten dimensions, written in a metric explicitly rescaled
with $e^{a\phi}$ with an arbitrary constant $a$
\bea
\label{1}
S = \frac{1}{\kappa^2}\int d^{10}x 
\sqrt{-g}e^{4\left( a+\frac{1}{3}\right)\phi}
\left[ R-9a \left( a+\frac{2}{3}\right)\partial\phi^2+\frac{3}{2}
e^{-2\left( a+\frac{4}{3}\right)\phi}H^2\right]
\nonumber\\
-\frac{T_2}{2}\int d^2\xi \left[
\sqrt{-\gamma}\gamma^{ij}\partial_iX^M
\partial_j X^Ne^{\left( b+\frac{4}{3}\right)\phi}g_{MN}
\right.
\nonumber\\
\left.
+2\varepsilon^{ij}
\partial_iX^M\partial_jX^NB_{MN}\right].
\eea
Here $a=-\frac{4}{3}$ gives us the usual (heterotic) string metric,
$a=-\frac{1}{3}$ the Einstein metric, and $a=\frac{2}{3}$ the type I string 
metric. 

For $b=a$ we have a heterotic string source, while
$b=a-2$ and, no $B_{MN}$ term in the string part of 
the action will
give us a type~I source. Since we only consider solutions
with the Yang-Mills fields identically zero, we omit the $\mbox{tr}{F^2}$
term in the action just like we do with all the fermionic terms. 

The variation of (\ref{1}) with respect to $g_{MN}$, $B_{MN}$, $\phi$,
$\gamma_{ij}$, and $X^M$, gives us the supergravity equations of motion, with
$\delta$-function sources at the location of the string, 
and the equations of motion for the
string degrees of freedom. 

To find a string solution, we split up the
coordinates ($M=0,1,\ldots 9$)
\be
\label{2}
x^M=(x^\mu,y^m)
\ee
where $\mu=0,1$ and $m=2,\ldots 9$, and make the ansatz
\be
\label{3}
ds^2=e^{2A}\eta_{\mu\nu}dx^\mu dx^\nu-e^{2B}\delta_{mn}dy^mdy^n,
\ee
\be
\label{4}
B_{\mu\nu}=\gamma\frac{\varepsilon_{\mu\nu}}{\sqrt{g_2}}e^C,
\ee
with $g_2 = -\mbox{det}g_{\mu\nu}$.
All other fields are put equal to zero, and the only coordinate dependance is 
on
$y=\sqrt{\delta_{mn}y^my^n}$. We need a numerical constant $\gamma$ in the
definition of $B_{\mu\nu}$ because of our nonstandard normalization.
The string coordinate $X^M(\xi)$ is split up in the same way as the
coordinates (\ref{2}), and we make the static gauge choice $X^\mu=\xi^\mu$,
and assume $Y^m=\mbox{constant}$, put to zero for simplicity.

The $\gamma_{ij}$ equation now immediately expresses $\gamma_{ij}$
as a function of the metric and the string coordinates.
To find the general solution to the remaining equations of motion
it turns out to be useful to make the field redifinitions
\be
\label{10}
X = 2A + 6B + 4\left(a+\frac{1}{3}\right)\phi,
\ee
\be
\label{11}
Y = 2A + \left(a-\frac{8}{3}\right)\phi,
\ee
\be
\label{12}
Z = 2A + \left(a+\frac{4}{3}\right)\phi.
\ee
The equations of motion are then equivalent to
\bea
\label{13}
e^{X-2A}\left[\frac{7}{6}(\nabla^2X + \frac{1}{2}X'^2) - \frac{1}{6}
(\nabla^2Y + X'Y' - \frac{1}{2}Y'^2) 
\right. 
\nonumber\\
\left.
- \frac{1}{2}(\nabla^2Z + X'Z'
- \frac{1}{2}Z'^2) + \gamma^2 e^{-2Z}\partial\left(e^{2A+C}\right)^2\right]
\nonumber\\
= -\kappa^2T_2 e^{Z-2A+(b-a)\phi}\delta^8(y),
\eea
\be
\label{14}
e^{X-2B}\left[X'' + \frac{13X'}{y} + X'^2\right] = 0,
\ee
\bea
\label{15}
e^{X-2B}\left[7\left(X''-\frac{1}{12}X'^2\right) + \frac{13}{12}Y'^2
+ \frac{13}{4}Z'^2 \right.
\nonumber\\
\left.
- 13 \gamma^2 e^{-2Z}\left(\partial e^{2A+C}\right)^2
\right] = 0,
\eea
\be
\label{16}
\gamma\frac{1}{y^7}\partial\left(y^7e^{X-2Z}\partial e^{2A+C}\right)
= \kappa^2T_2\delta^8(y),
\ee
\bea
\label{17}
e^X\left[\frac{14}{3}\left(a+\frac{1}{3}\right)(\nabla^2X+\frac{1}{2}X'^2)
-\frac{1}{6}\left(a-\frac{8}{3}\right)(\nabla^2Y + X'Y')
\right.
\nonumber\\
+ \frac{1}{3}\left(a+\frac{1}{3}\right)Y'^2  
-\frac{1}{2}\left(a+\frac{4}{3}\right)(\nabla^2Z+X'Z') +
\left(a+\frac{1}{3}\right)Z'^2 
\nonumber\\
\left.
- 2\gamma^2\left(a-\frac{2}{3}\right)
e^{-2Z}\left(\partial e^{2A+C}\right)^2\right]
\nonumber\\ 
= -\kappa^2T_2\left(b+\frac{4}{3}\right)e^{Z+(b-a)\phi}\delta^8(y),
\eea
\be
\label{18}
\partial e^{Z+(b-a)\phi} = 2\gamma\partial e^{2A+C}.
\ee

First we
now to solve all equations for $y>0$, and only afterwards 
we consider the singularity structure at $y=0$ to find a consistent
choice for the values of the integration constants.
In the following, a subscript zero will always denote
the value of the function in question at infinity, and $K,L \ldots$
are integration constants.
Choosing 
$\gamma = \frac{1}{2}$ (cf (\ref{18}) and \cite{T96}), and assuming $L\neq 0$
(solutions with a conserved charge), the supergraavity equations have the 
solution
\be
\label{19}
e^X = e^{X_0} + \frac{K}{y^{12}},
\ee
\be
\label{25}
e^XY' = \frac{M}{y^7},
\ee
\be
\label{26}
\frac{Z'}
{\left(L^2e^{2Z} - \frac{M^2}{3} - 7 \cdot 48K e^{X_0}\right)^{1/2}}
= \pm \frac{e^{-X}}{y^7},
\ee
\be
\label{261}
e^{2A+C}= \pm
\frac{\left(L^2e^{2Z} - \frac{M^2}{3} - 7 \cdot 48K e^{X_0}\right)^{1/2}}{L}
+\mbox{constant} .
\ee
Equations \ref{25} and \ref{26} can be integrated once more, and \ref{26} 
solved
for $e^{-Z}$ in terms of $y$, but the exact form depends on the values of the
integration constants, so we will not do this for the general case. The
explicit form for $K<0$
(for generic dimensions) can be found in a paper by L\"u et al. \cite{LPX}. 
One of the conditions for a supersymmetric solution (see below) is $X'=0$, so 
we immediately see that our nontrivial candidate to a generalization cannot
preserve supersymmetry. 

The equation of motion for the string source we have so 
far neglected gives us
\be
\label{20}
e^{2A+C} = e^{Z+(b-a)\phi} +\mbox{constant}.
\ee
The precise value of the constant is not very interesting since it does not
affect $H_{m\mu\nu}$.

As implied by Dabholkar et al. \cite{Dab95,DGHW96}, this type of solution 
really 
requires a compactification. This comes about as follows: Varying the 
supergravity part of our action with respect to some field, we obtain the
usual equations of motion, while the string part of the action gives terms of
the form $\int d^2\xi\delta^{10}(X^\mu(\xi)-x^\mu)\times \mbox{fields}$. 
With our ansatz
this becomes $\int d\tau\delta(\tau-x^0)\int_0^{2\pi}d\sigma\delta(\sigma-x^1)
\delta^8(y)$. The $\tau$-integration immediately gives a factor one,
while it is not at all obvious that this, which has always been assumed, 
should also be the case for the $\sigma$-part.
To obtain a one also here, all possible values of $x^1$ must lie within 
the interval of possible values of $\sigma$, $0\leq x^1 < 2\pi(\times R)$. 
The obvious interpretation is that $x^1$ is compactified, and that the source 
string
winds around it once. We can then also consider strings winding $n$ times 
around the compactified coordinate. Then $n$ different values of $\sigma$ 
correspond to
each $x^1$, and we obtain $\int_0^{2\pi}d\sigma\delta(\sigma - x^1)=n$.
The generalization to higher branes is obvious.

\section{The heterotic string solution} 

We will study the solution for $L\neq 0$ given in Section~2, 
and show that this is the one corresponding to the elementary string 
solution \cite{DGHR90}. By a full analysis of the zero modes, it can
be shown 
that this is indeed a heterotic string
\cite{Dab95,H195}. 

In order to satisfy equation \ref{20}
for $L \neq 0$, we must choose
\be
\label{29}
b=a
\ee
\be
\label{30}
K=-\frac{M^2e^{-X_0}}{7\cdot 144}
\ee
This is inserted into our equations,  which can then be integrated to
\be
\label{32}
Y =  \left\{\begin{array}{ll}Y_0 +
\frac{\sqrt{7}M}{|M|}\log \left|\frac{y^6-\frac{|M|e^{-X_0}}{\sqrt{7}\cdot 12}}
{y^6 +\frac{|M|e^{-X_0}}{\sqrt{7}\cdot 12}}\right| & M \neq 0
\\
Y_0 & M=0
\end{array}
\right.
\ee
and
\be
\label{34}
e^{-Z} = \left\{\begin{array}{ll} e^{-Z_0} -
\frac{\sqrt{7}L}{|M|}\log \left|\frac{y^6-\frac{|M|e^{-X_0}}{\sqrt{7}\cdot 12}}
{y^6 + \frac{|M|e^{-X_0}}{\sqrt{7}\cdot 12}}\right| & M \neq 0
\\
e^{-Z_0} + \frac{Le^{-X_0}}{6y^6} & M=0
\end{array}
\right. .
\ee
A study of the singularities at $y=0$ yields
\be
\label{35}
e^XY' = M\Omega_7 f',
\ee
\be
\label{36}
e^X\partial e^{-Z} = -L\Omega_7 f',
\ee
where $\nabla^2 f = \delta^8(y)$, and $\Omega_7$ is the volume of the 
seven-sphere. The nonvanishing parts of the equations of motion at $y=0$
are then
\be
\label{37}
e^{-2A}\left[-\frac{1}{6}M - \frac{1}{2}Le^Z\right]\Omega_7\delta^8(y)
= -\kappa^2T_2 e^{Z-2A}\delta^8(y),
\ee
\be
\label{38}
L\Omega_7\delta^8(y) = 2\kappa^2 T_2 \delta^8(y),
\ee
\bea
\label{39}
\left[-\frac{1}{6}\left(a-\frac{8}{3}\right)M - \frac{1}{2}\left(a+\frac{4}{3}
\right)Le^Z\right]\Omega_7\delta^8(y) 
\nonumber\\
= - \kappa^2T_2\left(a+\frac{4}{3}\right)
e^Z \delta^8(y),
\eea
so the constants must take the values 
\be
\label{40}
L = \frac{2\kappa^2T_2}{\Omega_7},
\ee
\be
\label{41}
M = 0.
\ee

The remaining integration constants are the values of the fields at infinity,
$X_0$, $Y_0$ and $Z_0$, which can be rewritten in terms of $A_0$, $B_0$ and 
$\phi_0$. The first two of these can be removed
by constant rescaling of the coordinates, so we are left with only $\phi_0$,
which is the vacuum expectation value of the dilaton field.
This is exactly the standard elementary string solution \cite{DGHR90,DKL94},
written 
in arbitrary metric.
We have a preserved N\oe ther charge 
\be
\label{42}
e = \sqrt{2}\kappa T_2.
\ee
The factor 6 reflects our different normalization of $H$ with respect to
e.g. Duff et al. \cite{DKL94}. 
The mass per unit string length is
\be
\label{43}
{\cal M}_2 = 
 2\left(a+\frac{5}{6}\right)T_2 e^{\left(a+\frac{4}{3}\right)\phi_0},
\ee
and we can see explicitly that ${\cal M}_2$  scales with $g_{MN}$ as 
it should \cite{H295}. 

For the $n$-winding state $T_2$ is just replaced by $nT_2$ everywhere, which 
gives exactly the mass per unit length of the $(0,n)$ winding states of 
Dabholkar et al. \cite{DGHW96}.

The conditions that the solution preserve half the supersymmetry can 
be obtained just like in the papers quoted above. We find
\be
\label{47}
X' = 0,
\ee
\be
\label{48}
Y'= 0,
\ee
\be
\label{49}
\partial e^{Z} = \partial e^{2A+C}.
\ee
These equations are all satisfied here as we know they should be.

Provided $a+\frac{4}{3} > 0$ 
the solution we have found can also be interpreted as a solitonic
string solution of the dual version of ten-dimensional supergravity.
If we require $e^{-Z}$ sufficiently well-behaved this solution is again
unique.

\section{The type~I string solution}
So far, we excluded
$L=0$. In this case the lhs of equation \ref{16} has no singularity at
$y=0$ and hence we cannot have a source at the rhs. We then have to redo
the analysis putting $\partial e^{2A+C}=0$ in (\ref{13}), (\ref{15}), 
(\ref{17}) and
(\ref{18}), and removing the rhs of (\ref{16}). 

We first solve equations \ref{13}-\ref{17} for $y>0$. Equation \ref{14}
gives the same solution as before for $e^X$, and the remaining equations 
have the solution
(for $K$ different from zero)
\be
\label{54}
Y = Y_0 + \frac{Me^{-X_0/2}}{12\sqrt{-K}}\log \left|
\frac{y^6-\sqrt{-K}e^{-X_0/2}}{y^6+\sqrt{-K}e^{-X_0/2}}\right|  ,
\ee
\be
\label{55}
Z = Z_0 + \frac{Ne^{-X_0/2}}{12\sqrt{-K}}\log \left|
\frac{y^6-\sqrt{-K}e^{-X_0/2}}{y^6+\sqrt{-K}e^{-X_0/2}}\right| ,
\ee
\be
\label{56}
K = - \frac{M^2 + 3N^2}{7\cdot 144}e^{-X_0}.
\ee
At $y=0$ we now have 
\be
\label{57}
\left[-\frac{1}{6}M -\frac{1}{2}N\right] \Omega_7\delta^8(y)
= - \kappa^2T_2e^{Z-2A+(b-a)\phi}\delta^8(y),
\ee
\bea 
\label{58}
\left[-\frac{1}{6}\left(a-\frac{8}{3}\right)M - \frac{1}{2}\left(a+\frac{4}{3}
\right)N\right]\Omega_7\delta^8(y) 
\nonumber\\
= -\kappa^2T_2\left(b+\frac{4}{3}\right)
e^{Z+(b-a)\phi}\delta^8(y).
\eea
These equations have the solutions
\be
\label{59}
M = -\frac{3}{2}\frac{\kappa^2T_2}{\Omega_7}(b-a)e^{(Z+(b-a)\phi)(0)},
\ee
\be
\label{60}
N =  \frac{1}{2}\frac{\kappa^2T_2}{\Omega_7}(b-a+4)e^{(Z+(b-a)\phi)(0)}.
\ee
This is in contradiction with (\ref{18}), which
requires $(b-a+4)^2+3(b-a)^2=0$. However, equation \ref{18} is the
equation of motion for the string source. 
It can be interpreted as a
no-force condition, stating that the graviton contribution to the force
between two parallell source strings is cancelled by the ``axion'' 
contribution. In the present case we have no axion contribution (to the 
lowest order in $\alpha'$ at least). Furthermore, this solution does not
preserve supersymmetry, nor does it have a conserved N\oe ther charge,
so there is no reason to expect it to be stable. It is hence just the 
configuration around a source term corresponding to an 
infinitely long string put in by hand. 
There is no reason 
to expect that such an  unstable test source should satisfy dynamical
equations of motion, or that there should be no force between two parallell 
unstable strings, so we just drop
equation \ref{18}.

The reason we are still interested in this solution is that we 
are looking
for just such a thing as the missing type~I ``soliton", \cite{H295}. 
The mass per unit string length for our solution is
\be
\label{61}
{\cal M}_2 = \frac{T_2}{2}\left[3\left(a+\frac{1}{6}\right)(b-a) +
\left(a+\frac{5}{6}\right)(b-a+4)\right]e^{\left(b+\frac{4}{3}\right)\phi_0}
\ee
If we choose $b=a-2$, corresponding to a type~I string source we obtain 
\be
\label{62}
{\cal M}_2 = -2\left(a-\frac{1}{2}\right)e^{\left(a-\frac{2}{3}\right)\phi_0}.
\ee
This is indeed the correct scaling behaviour for a type~I soliton, 
since $a=\frac{2}{3}$ in the type~I metric. The fact that there is no
$B_{MN}$ term in the source is also consistent with the type~I interpretation,
since $B_{MN}$ is here a Ramond-Ramond state corresponding to 
a sigma model term $\gamma^{MNP}H_{MNP}$ sandwiched between two spin fields
\cite{CLNY88}, and such a term should vanish in
our ansatz.

For this ``incomplete type~I solution", comparable to heterotic solutions
where we explicitly choose not to satisfy \ref{18}, there is no alternative
soliton interpretation, nor does the analysis of the zero-modes produce 
anything useful.

\section{Final remarks}
We have found the general solution to the equations of motion for 
ten-dimension\-al
 supergravity coupled to a three-form field strengh given the standard
string ansatz, and we show exactly when this solution is restricted to 
the fundamental
(heterotic) string solution. Assuming that there is no conserved
charge, we also saw that there is a possible interpretation as a type~I
string solution.

The calculation is easy to generalize to arbitrary $p-1$-brane solutions
in an arbitrary dimension, see L\"u et al. \cite{LPX}, and Tollst\'en 
\cite{T96}.
The first authors limit themself to the solution for $K<0$, and nonzero $L$,
for which case, however, they give the explicit solution, which is interpreted
as a non-extremal black $p-1$-brane. Whether there exist physical 
interpretations 
for other cases remains unclear.

\section*{Acknowledgments}
I am grateful for discussions with Ansar Fayyazuddin.

\newcommand{\Journal}[4]{{#1} {\bf #2}, #3 (#4)}
\newcommand{\NCA}{\em Nuovo Cimento}
\newcommand{\NIM}{\em Nucl. Instrum. Methods}
\newcommand{\NIMA}{{\em Nucl. Instrum. Methods} A}
\newcommand{\NPB}{{\em Nucl. Phys.} B}
\newcommand{\PLB}{{\em Phys. Lett.}  B}
\newcommand{\PRL}{\em Phys. Rev. Lett.}
\newcommand{\PRD}{{\em Phys. Rev.} D}
\newcommand{\ZPC}{{\em Z. Phys.} C}
\newcommand{\PRep}{\em Phys. Rept}
\newcommand{\CQG}{\em Class. Quantum Grav.}
\newcommand{\MPLA}{{\em Mod. Phys. Lett.} A}

\end{document}